\documentstyle[12pt,world_sci]{article}

\newcommand{\beqn}{\begin{eqnarray}}
\newcommand{\eeqn}{\end{eqnarray}}

\def\spose#1{\hbox to 0pt{#1\hss}}
\def\lsim{\mathrel{\spose{\lower 3pt\hbox{$\mathchar"218$}}
     \raise 2.0pt\hbox{$\mathchar"13C$}}}
\def\gsim{\mathrel{\spose{\lower 3pt\hbox{$\mathchar"218$}}
     \raise 2.0pt\hbox{$\mathchar"13E$}}}
\def\simpropto{\mathrel{\spose{\lower 3pt\hbox{$\mathchar"218$}}
     \raise 2.0pt\hbox{$\propto$}}}

\def\beq{\begin{equation}}
\def\eeq{\end{equation}}
\def\etal{ {\it et al.} }

\def\PRL#1#2#3{{\sl Phys. Rev. Lett.} {\bf #1}, #2 (#3)}
\def\PRD#1#2#3{{\sl Phys. Rev.} {\bf D#1}, #2 (#3)}
\def\PLB#1#2#3{{\sl Phys. Lett.} {\bf B#1}, #2 (#3)}

\def\NPB#1#2#3{{\sl Nucl. Phys.} {\bf B#1}, #2 (#3)}

\catcode`\@=11
\long\def\@makefntext#1{
\protect\noindent \hbox to 3.2pt {\hskip-.9pt  
$^{{\ninerm\@thefnmark}}$\hfil}#1\hfill}                

\def\@makefnmark{\hbox to 0pt{$^{\@thefnmark}$\hss}}  
        
\def\ps@myheadings{\let\@mkboth\@gobbletwo
\def\@oddhead{\hbox{}
\rightmark\hfil\ninerm\thepage}   
\def\@oddfoot{}\def\@evenhead{\ninerm\thepage\hfil
\leftmark\hbox{}}\def\@evenfoot{}
\def\sectionmark##1{}\def\subsectionmark##1{}}

\renewcommand{\thefootnote}{\fnsymbol{footnote}}

\newcounter{sectionc}\newcounter{subsectionc}\newcounter{subsubsectionc}
\renewcommand{\section}[1] {\vspace*{0.6cm}\addtocounter{sectionc}{1} 
\setcounter{subsectionc}{0}\setcounter{subsubsectionc}{0}\noindent 
        {\normalsize\bf\thesectionc. #1}\par\vspace*{0.4cm}}
\renewcommand{\subsection}[1] {\vspace*{0.6cm}\addtocounter{subsectionc}{1} 
        \setcounter{subsubsectionc}{0}\noindent 
        {\normalsize\it\thesectionc.\thesubsectionc. #1}\par\vspace*{0.4cm}}
\renewcommand{\subsubsection}[1]
{\vspace*{0.6cm}\addtocounter{subsubsectionc}{1}
        \noindent {\normalsize\rm\thesectionc.\thesubsectionc.\thesubsubsectionc. 
        #1}\par\vspace*{0.4cm}}

\newcounter{appendixc}
\newcounter{subappendixc}[appendixc]
\newcounter{subsubappendixc}[subappendixc]

\renewcommand{\appendix}[1] {\vspace*{0.6cm}
        \refstepcounter{appendixc}
        \setcounter{figure}{0}
        \setcounter{table}{0}
        \setcounter{equation}{0}
        \renewcommand{\thefigure}{\Alph{appendixc}.\arabic{figure}}
        \renewcommand{\thetable}{\Alph{appendixc}.\arabic{table}}
        \renewcommand{\theappendixc}{\Alph{appendixc}}
        \renewcommand{\theequation}{\Alph{appendixc}.\arabic{equation}}
        \noindent{\bf Appendix \theappendixc #1}\par\vspace*{0.4cm}}

\def\abstracts#1{{
        \centering{\begin{minipage}{12.2truecm}\footnotesize\baselineskip=12pt\noindent
        \centerline{\footnotesize ABSTRACT}\vspace*{0.3cm}
        \parindent=0pt #1
        \end{minipage}}\par}} 

\renewenvironment{thebibliography}[1]
        {\begin{list}{\arabic{enumi}.}
        {\usecounter{enumi}\setlength{\parsep}{0pt}
\setlength{\leftmargin 1.25cm}{\rightmargin 0pt}
         \setlength{\itemsep}{0pt} \settowidth
        {\labelwidth}{#1.}\sloppy}}{\end{list}}

\newcounter{itemlistc}
\newcounter{romanlistc}
\newcounter{alphlistc}
\newcounter{arabiclistc}

\newcommand{\fcaption}[1]{
        \refstepcounter{figure}
        \setbox\@tempboxa = \hbox{\footnotesize Fig.~\thefigure. #1}
        \ifdim \wd\@tempboxa > 6in
           {\begin{center}
        \parbox{6in}{\footnotesize\baselineskip=12pt Fig.~\thefigure. #1}
            \end{center}}
        \else
             {\begin{center}
             {\footnotesize Fig.~\thefigure. #1}
              \end{center}}
        \fi}

\newcommand{\tcaption}[1]{
        \refstepcounter{table}
        \setbox\@tempboxa = \hbox{\footnotesize Table~\thetable. #1}
        \ifdim \wd\@tempboxa > 6in
           {\begin{center}
        \parbox{6in}{\footnotesize\baselineskip=12pt Table~\thetable. #1}
            \end{center}}
        \else
             {\begin{center}
             {\footnotesize Table~\thetable. #1}
              \end{center}}
        \fi}

\def\@citex[#1]#2{\if@filesw\immediate\write\@auxout
        {\string\citation{#2}}\fi
\def\@citea{}\@cite{\@for\@citeb:=#2\do
        {\@citea\def\@citea{,}\@ifundefined
        {b@\@citeb}{{\bf ?}\@warning
        {Citation `\@citeb' on page \thepage \space undefined}}
        {\csname b@\@citeb\endcsname}}}{#1}}

\newif\if@cghi
\def\cite{\@cghitrue\@ifnextchar [{\@tempswatrue
        \@citex}{\@tempswafalse\@citex[]}}
\def\citelow{\@cghifalse\@ifnextchar [{\@tempswatrue
        \@citex}{\@tempswafalse\@citex[]}}
\def\@cite#1#2{{$\null^{#1}$\if@tempswa\typeout
        {IJCGA warning: optional citation argument 
        ignored: `#2'} \fi}}

 1
 1
 1

\font\ninerm=cmr9

\begin{document}
\begin{flushright}
MPI/PhT/96--104\\
UCDPHY-96-27\\
September 1996
\end{flushright}

\centerline{}
\vskip3.2cm

\centerline{\normalsize\bf NEUTRINO PROPERTIES AND SUSY WITHOUT $R$-PARITY
\footnote{
to be published in the Preceedings of the VIIIth Rencontres de Blois:
   "Neutrinos, Dark Matter and the Universe",
    Blois, France (June 8-12, 1996)}
}


\centerline{}
\centerline{}
\centerline{\footnotesize Ralf Hempfling}
\baselineskip=13pt
\centerline{\footnotesize\it Max-Planck-Institut f\"ur Physik,
Werner-Heisenberg-Institut,}
\baselineskip=12pt
\centerline{\footnotesize\it F\"ohringer Ring 6, 80805 Munich, Germany}
\centerline{\footnotesize\it and}
\centerline{\footnotesize\it 
Univ. of California at Davis, Dept. of Physics, Davis, CA 95616}
\centerline{\footnotesize E-mail: hempf@bethe.ucdvis.edu}
\vspace*{0.3cm}

\vspace*{8.2cm}
\abstracts{
In supersymmetric models without $R$-parity
neutrinos naturally become massive and mix with each other.
We explore the predictions of a very restricted 
model with only three free parameters
and find that this model naturally yields masses and mixing angles
compatible with experimental results from solar and atmospheric neutrino
experiments.
Furthermore, there is a tiny region
in parameter space where the solution to the solar neutrino puzzle
is compatible with either the LSND result or
the existence of significant hot dark matter neutrinos.
}
\clearpage 
\vspace*{0.6cm}
\normalsize\baselineskip=15pt
\setcounter{footnote}{0}
\renewcommand{\thefootnote}{\alph{footnote}}

\begin{figure}
\vspace*{13pt}
\vspace*{6.7truein}      
\includegraphics{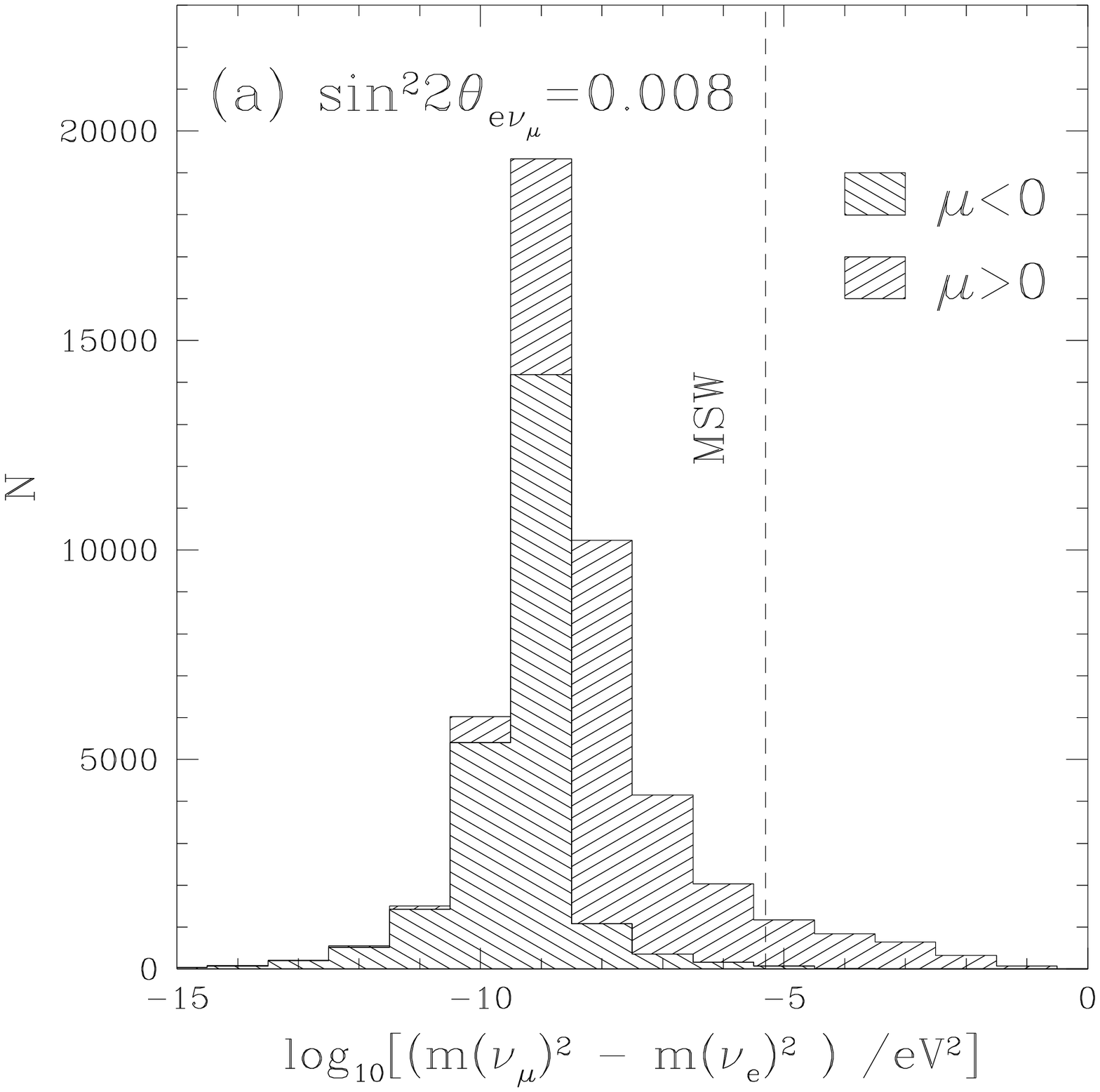}
\includegraphics{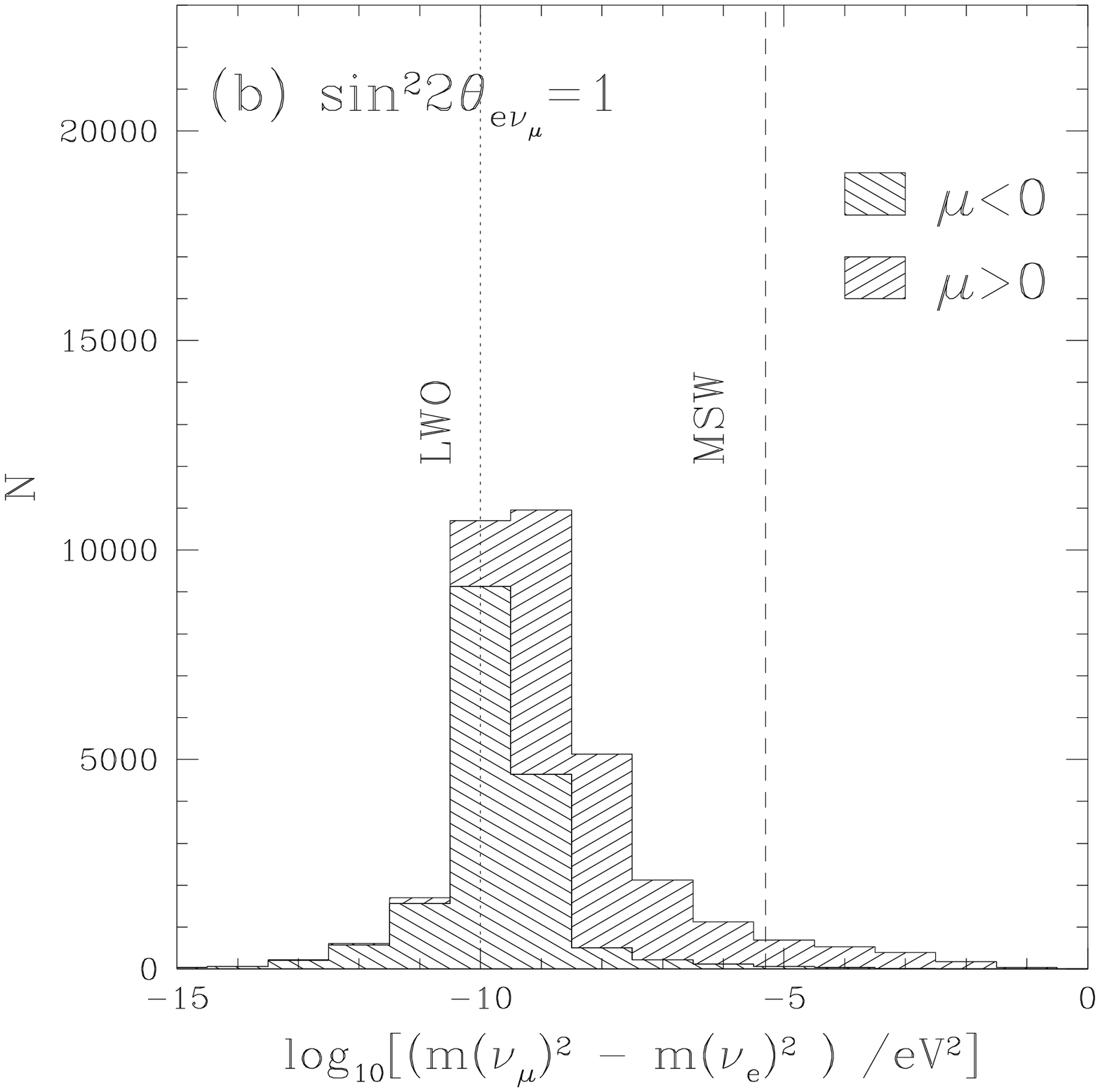}
\includegraphics{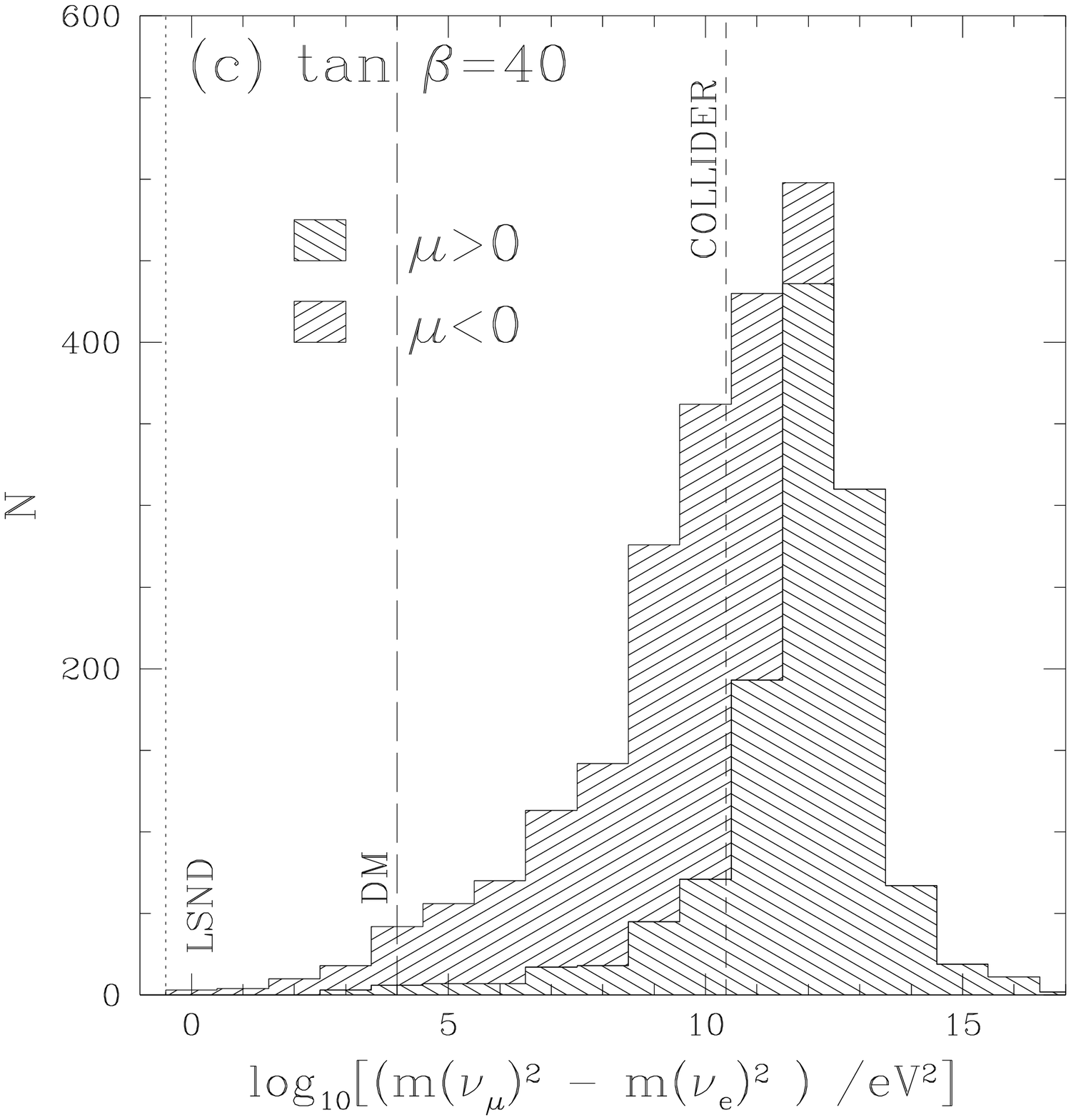}
\caption{Histogram
of the number of models that yield a particular prediction
for $m_{\nu_{\mu}}^2- m_{\nu_{e}}^2$ 
assuming (a) small angle and (b) large angle solution to solar
neutrino problem. In (c) we 
solve the solar neutrino problem via small angle $e$--$\tau$ 
oscillations and check whether this is compatible with
the LSND result.
}
\label{fig}
\end{figure}

In the Standard Model of elementary particles (SM)
both lepton number ($L$) and baryon number ($B$) are
conserved due to an accidental symmetry,
{\sl i.e.} there is no renormalizable, gauge-invariant
term that would break the symmetry.
In the minimal supersymmetric extension of the SM (MSSM)
the situation is different. Due to a the variety
of scalar partners the MSSM allows for a host of new 
interactions many of which violate $B$ or $L$.

Since neither $B$ nor $L$ violation has been
observed in present collider experiments
these couplings are constrained from above.
More constraints arise from neutrino
physics
or cosmology.
Thus, all lepton and baryon number violating
interaction are often eliminated by imposing
a discrete, multiplicative symmetry called
$R$-parity,\cite{r-parity}
$R_p \equiv (-1)^{2S+3B+L}$, where $S$ is the spin.
One very attractive feature of 
$R_p$ conserving models is that
the lightest supersymmetric particle (LSP)
is  stable and a good cold dark matter candidate.\cite{cdm}

However, while the existence of a dark matter candidate
is a very desirable prediction, it does not prove
$R_p$ conservation and 
one should consider more general models.
Here, we will investigate
the scenario where $R_p$ is broken explicitly via
the terms\cite{suzuki} $W = \mu_i L_i H$,
where $H$   is the Higgs coupling to up-type fermions
and $L_i$ ($i = 1,2,3$) are the left-handed lepton doublets.
Clearly, these Higgs-lepton mixing terms violate
lepton-number. As a result, majorana masses will be generated for
one neutrino at tree-level and for the remaining
two neutrinos at the one-loop level.
These masses were calculated in the frame-work of minimal supergravity
in ref.~\citenum{npb} and the numerical results will be
briefly summarized here.

There are three $R_P$ violating parameters which can be used to fix 
1) the tree-level neutrino mass,
2) the $\mu$--$\tau$ mixing angle and
3) the $e$--$\mu$ mixing angle.
The question of whether e.g. the solar\cite{solarn}
and the atmospheric\cite{atmosphericn} neutrino puzzle
can be solved simultaneously depends on the prediction of
$m_{\nu_\mu}^2-m_{\nu_e}^2$.
In fig.~1 we have scanned the entire SUSY parameter space
consisting of the Higgsino (gaugino) mass parameter,
$\mu$ ($m_{1/2}$), the trilinear scalar interaction parameter $A_0$,
and the ratio of Higgs VEVs, $\tan\beta$. The universal 
scalar mass parameter $m_0$ is fixed by minimizing the potential.
Plotted is the number of models yielding a particular prediction for
$m_{\nu_\mu}^2-m_{\nu_e}^2$ for
(a) sin$^2 2 \theta_{e \nu_\mu} = 0.008$ and
(b) sin$^2 2 \theta_{e \nu_\mu} = 1$.
We fix $m_{\nu_\tau}=0.1$~eV and 
sin$^2 2 \theta_{\mu \nu_\tau} = 1$ in order to solve the 
atmospheric neutrino problem.
We see that both
long wave-length oscillation (LWO)\cite{lwo}
($m_{\nu_\mu}^2-m_{\nu_e}^2=10^{-10}$~eV$^2$)
and MSW effect\cite{msw-effect}
($m_{\nu_\mu}^2-m_{\nu_e}^2=10^{-5}$~eV$^2$)
can be accommodated.
In fig.~1(c) we solve the solar neutrino problem via $e$--$\tau$
oscillations and we fix sin$^2 2 \theta_{e \nu_\mu} = 0.004$
in order to accommodate the LSND result.\cite{lsnd}
We see that most models are already ruled out
by collider constraints and even more by dark matter (DM) constraints.
However, a very small (but non-zero) number of models
yields a prediction compatible with the LSND
result (the dotted line is lower limit of LSND).

\noindent{\bf Acknowledgements}
This work was supported in parts by the DOE under
Grants No. DE-FG03-91-ER40674 and by the
Davis Institute for High Energy Physics.


\vskip0.3cm
\noindent{\bf References}
\begin{thebibliography}{9}

\bibitem{r-parity}
N. Sakai and T. Yanagida, \NPB{197}{533}{1982}.

\bibitem{cdm} J. Ellis \etal,
\NPB{238}{453}{1984}.

\bibitem{suzuki} L.J. Hall and M. Suzuki, \NPB{231}{419}{1984}.

\bibitem{npb} R. Hempfling, MPI-PhT/95-59, hep-ph/9511288,
 {\sl Nucl. Phys.} {\bf B}, to appear.

\bibitem{solarn}
P. Anselmann \etal, \PLB{327}{234}{1994}.

\bibitem{atmosphericn}
Y. Fukunda \etal, \PLB{335}{237}{1994}.

\bibitem{lwo}
V. Gribov and B. Pontecorvo, \PLB{28}{493}{1969};
V. Barger, R.J.N. Phillips and K. Whisnant, \PRD{24}{538}{1981};
\PRL{69}{3135}{1992}.

\bibitem{msw-effect} L. Wolfenstein, \PRD{17}{2369}{1978};
  {\bf 20}, 2634 (1979);
  S.P. Mickheyev and A. Yu Smirnov, {\sl Yad. Fiz.}
 {\bf 42}, 1441 (1985) [{\sl Sov. J. Nucl. Phys.} {\bf 42}, 913 (1986)].

\bibitem{lsnd} 
C. Athanassopoulos \etal, \PRD{75}{2650}{1995}; LA-UR-96-1326.

\end{thebibliography}
\end{document}